\documentclass[twocolumn,showpacs,preprintnumbers,amsmath,amssymb]{revtex4}
\usepackage{graphicx}% Include figure files
\usepackage{dcolumn}% Align table columns on decimal point
\usepackage{bm}% bold math

\begin{document}

%\preprint{APS/123-QED}

\title{Quantum Cryptography: from Theory to Practice}% Force line breaks with \\

\author{Hoi-Kwong Lo and Norbert L\"{u}tkenhaus}
\affiliation{Center for Quantum Information and Quantum Control,
Department of Physics and Department of Electrical \& Computer
Engineering, University of Toronto, Toronto, Ontario, M5S 3G4,
Canada \\
and \\
Institute for Quantum Computing \& Department of Physics and Astronomy, University of Waterloo,
200 University Ave. W.
N2L 3G1
}%

\date{\today}% It is always \today, today,

\begin{abstract}

Quantum cryptography can, in principle, provide unconditional security
guaranteed by the law of physics only. Here, we survey the theory and practice
of the subject and highlight some recent developments.

\end{abstract}

\pacs{}% PACS, the Physics and Astronomy
\maketitle

%\section{Introduction}

"The human desire to keep secrets is almost as old as writing itself.'' \cite{welsh88a}
With the advent in electronic commerce,
the importance of secure communications via encryption is growing.
Each time when we go on-line for internet banking, we should be
concerned with communication security.
Indeed, methods of secret communications were used in many ancient civilizations
including Mesopotamia, Eqypt, India and China.
Legends say that Julius Caesar used a simple substitution cipher in his
correspondences. Each letter is replaced by a letter that
followed it three places alphabetically. For instance, the word LOW is
replaced by ORZ because the first letter "L" in LOW is replaced by
$L \to M \to N \to O$, etc. Regardless of the size of the shift, the
illustrated method is still called Caesar's cipher.

Let us introduce the general problem of secure
communication.
Suppose a sender, Alice, would like to send a message to a receiver, Bob.
An eavesdropper, Eve, would like to
learn about the message. How can Alice prevent Eve from
learning her message?

A standard method is encryption. Alice uses her encryption key (some secret
information) to transform her message (a plaintext) into something (a ciphertext)
that is unintelligible to Eve and sends the ciphertext through a communication
channel. Bob, with his decryption key, recovers Alice's message from the
ciphertext. For instance, in Caesar's cipher, the key takes a
value between 1 and 26, which denotes the size of the shift.

Since encryption machines may be captured, in modern cryptography,
it is standard to assume that the encryption method is known
and the security of the message lies on the security of the key.
For instance, we assume that Eve knows that Caesar's cipher is being
used, but she does not {\em a priori} know the value of the key.
Note that Caesar's cipher is not that secure because the number of
all possible
key values is so small (only 26) that Eve can easily try
all possible values.

Throughout the history of cryptography, many ciphers have been
invented and believed for a while to be unbreakable. Almost
all have subsequently been broken, with disastrous consequences to
the unsuspecting users. For instance, in the Second World War, the Allies' breaking
of the German Enigma code contributed greatly to the ultimate victory of the Allies.
The first lesson in cryptography is: never under-estimate the ingenuity
and efforts that your enemies are willing to spend on breaking your codes.

Unbreakable codes do exist in conventional cryptography.
They are called one-time pads and were invented by Gilbert Vernam in 1918:
A message is first
converted into a binary form (i.e., a string consisting of 0's and 1's) by
a publicly known method.
The key is another sequence of 0's and 1's of the same length.
For encryption, Alice combines each bit of the message with
the respective bit of the key
by using addition modulo 2 to generate the ciphertext.
For decryption, Bob combines each bit of the ciphertext with
the respective bit of the key by using addition modulo 2 to
generate the plaintext.

For one-time pad to be secure, it is important that a key is
never re-used. This is why it is called a one-time pad.
Why re-using a key would make the scheme insecure?
Suppose the same key, $k$, is used to encrypt two different
messages, $m_1 $ and $m_2$. Then, the cipher-texts, $c_1 = m_1 \oplus k$
and $c_2 = m_2 \oplus k$, are transmitted in public.
An eavesdropper can simply take the addition modulo 2 of the
two cipher-texts to obtain
$c_1 \oplus c_2= m_1 \oplus k \oplus m_2 \oplus k =m_1 \oplus m_2  $.
But, this allows Eve to learn some non-trivial information,
namely the parity, about the
two original messages.

\section{Key Distribution Problem}

The one-time pad has a serious drawback: it pre-supposes that
Alice and Bob share a random string of secret, a key, before
the actual transmission of the message.
So, the introduction of the one-time pad shifts the problem of
secure communication to the
problem of secure key distribution.
This is called the key distribution problem.
In top secret applications,
key distribution is often done by trusted couriers.
Recall that in one-time pad,
a key must be as long as a message.
Sending long keys by trusted couriers is clearly rather inconvenient.

All conventional (classical) key distribution schemes are
fundamentally insecure because there is nothing to prevent an eavesdropper from
making a copy of the key during the key distribution process.
Indeed, trusted couriers could be bribed or compromised.
So, the users can never be sure about the security of a key.

How may one solve the key distribution problem?
Around 1970s, mathematicians invented public key cryptography.
In public key cryptography, there are two different sets of keys, the encryption
key and the decryption key. The encryption key can be broadcast in public
(e.g., published in a phone book) whereas the decryption key has to be
kept secret. Public key cryptography allows two parties who have never
met before to communicate securely.

Unfortunately, the security of public key encryption schemes is often
based on unproven computational assumptions.
For instance, the security of standard RSA encryption
scheme is based on the presumed hardness of factoring
a large composite number. Such an assumption may be
broken by unanticipated advances in algorithms and hardware.
For instance, in 1994 Peter Shor, then at AT\&T,
found an efficient quantum algorithm for factoring. \cite{shor97a}
Therefore, "if a quantum computer is ever built,
much of conventional cryptography will fall apart!"
(Gilles Brassard).

You may think, "since we do not have a quantum computer yet,
perhaps, we should not worry about this problem until
a quantum computer has been built." Not so.
For instance, Canada has kept
census information secret for 92 years on average.
An eavesdropper may save messages
sent by you in 2007 and try to
decrypt them in 2099.
And, who knows whether we will have a quantum computer by 2099?

\section{quantum key distribution}

It is fortunate that quantum mechanics can also
come to the rescue. Unlike conventional cryptography,
the Holy Grail of quantum cryptography (code-making)
is unconditional security, that is to say,
security that is based on the fundamental law of quantum
mechanics, namely that information gain generally implies
disturbance on quantum states.

How does quantum key distribution work?
Intuitively, if an eavesdropper attempts to learn
information about some signals sent through a quantum channel,
she will have to perform some sort of measurement on the signals.
Now, a measurement will generally disturb the state of those signals.
Alice and Bob can catch an eavesdropper by searching for traces of
this disturbance. The absence of disturbance assures Alice and Bob
that Eve almost surely does not have any information about the
transmitted quantum signals.

\section{BB84 protocol: The ideal case}

The best-known quantum key distribution (QKD) protocol (BB84) was published
by Bennett and Brassard in 1984 \cite{bennett84a}, while its idea goes back to
Wiesner. \cite{wiesner83a}
The basic tool are a quantum channel connecting Alice and Bob and
a public classical channel, where Eve is allowed to
listen passively, but not allowed to change the transmitted message.
For the quantum channel, we use four signal states. For simplicity,
let us for now regard the signals as realized by single photons
in the polarization degree of freedom. Consider two sets of orthogonal
signals, one formed by a horizontal and a vertical polarized photon,
and the other formed by a 45-degree and 135-degree polarized photon.
These four polarized states are non-orthogonal. The overlap probability
between signals from two different sets is one half.
Bob has two measurement devices at his hand, one in the rectilinear
(i.e., vertical/horizontal) basis and one in the diagonal (i.e.,
45-degree/135-degree ) basis.
Notice that Bob's two measurements do not commute.

\begin{figure}[htb]
\includegraphics[width=6cm]{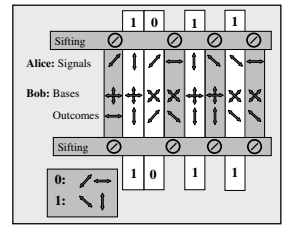}%
\caption{\label{fig:BB84} Schematics of the BB84 protocol. \cite{bennett84a}}
\end{figure}

The procedure of BB84 is as follows. See Figure \ref{fig:BB84}.

1. Phase I (Quantum Communication Phase)

(a) Alice sends a sequence of signals, each randomly chosen from one of
the above four polarizations.

(b) For each signal, Bob randomly chooses one of the two measurement devices
to perform a measurement.

(c) Bob confirms that he has received and measured all signals.

2. Phase II (Public Discussion Phase)

(a) Alice and Bob announce their polarization bases for each signal.
They discard all events where they use different bases for a signal.

(b) Alice randomly chooses a fraction, $p$, of all remaining events
as test events. For those test events,
she transmits the positions and the corresponding
polarization data to Bob.
Bob compares his polarization data with those of Alice and tells Alice
whether their polarization data for the test events agree.

(c) In case of agreement, Alice and Bob convert the polarization data
of the remaining set of events into
binary form, e.g., they call all horizontal and 45-degree signals
a "0" and all vertical and 135-degree signals a "1".
Such a generated binary string is now their secret key.

The first phase of the protocol uses signals and measurements via the quantum
channel. Alice keeps a classical record of the signal states she sent.
Similarly, Bob keeps a classical record of the measurement devices he
has chosen together with his measurement outcomes.

The second phase of the protocol uses a public classical channel.
An eavesdropper may try to break the scheme by launching a man-in-the-middle
attack where she impersonates as Alice to Bob and impersonates as Bob to
Alice. To prevent this attack, Alice and Bob should
authenticate the data sent in their classical channel.
Fortunately, efficient classical authentication methods exist.
To authenticate an $M$-bit message, Alice and Bob only need to consume
a key of order $\log M$ bits. In summary, starting from a short
pre-shared key, Alice and Bob can generate a long secure key by
using quantum key distribution. They can keep a small portion of it for
authentication in a subsequent round of
quantum key distribution and use the rest as a key for one-time pads.
The fact that there is no degradation of security by using this new secure
key is called composability and has been proven in
\cite{benor05a}, provided that one uses a proper definition of
security. See, e.g. \cite{benor05a,renner05a}.
As a result, we should strictly speak of QKD as
quantum key growing.

The BB84 protocol is only one example of
a QKD protocol.
Actually, there are many QKD protocols,
as nearly any set of non-orthogonal
signal states together with a set of non-commuting measurement devices will
allow secure QKD. \cite{gisin02a} These protocols differ, however,
in their symmetry that
simplifies the security analysis, in the ease of their experimental
realization and in their tolerance to channel noise and loss.
Independent of Bennett and Brassard's work,
Ekert proposed a QKD protocol (Ekert 91) based on Bell's inequalities \cite{ekert91}.
In 1992, Bennett proposed a simple protocol (B92)
\cite{bennett92b}
that involves only
two non-orthogonal states. A protocol of particularly high symmetry is the
six-state protocol.

\section{BB84 protocol in a noisy envirnoment}

The idealized BB84 protocol described above will not work in any practical
realizations.
Even when there are no eavesdropping activities, any real quantum channel
is necessarily noisy due to, for instance, some misalignment in a
quantum channel. As a result, Alice and Bob will generally
find a finite amount of
disturbance in their test signals.
Since Alice and Bob can never be sure about the origin of
the disturbance, as conservative cryptographers, we should assume that Eve
has full control of the channel. Therefore, we are faced with two problems.
First, the polarization data of Alice may be different from those of Bob.
This means that their raw keys are different. Second, Eve might have
some partial information on those raw keys.

In order to address these two problems, Alice and Bob have to perform
classical post-processing of their raw keys. First,
Alice and Bob may perform error correction to correct any error
in the raw key. Now, they share a reconciled key on which
Eve may have partial information.
Second, they may perform so-called privacy amplification.
That is to say, they apply a function on their reconciled key to
map it into a final key, which is shorter, but is supposed to be
almost perfectly secure.

Proving the security of QKD in a noisy setting was a very hard problem.
This is because instead of attacking Alice's signals individually,
Eve may conduct a joint attack. In the most general
attack, Eve may couple all the signals received from Alice with her probe and evolve
the combined system by some unitary transformation and then send parts of
her systems to Bob, keeping the rest in her quantum memory.
She then listens to all the public discussion between Alice and Bob.
Some time in the future, Eve may perform some measurement on her system to
try to extract some information about the key. A priori, it is very hard to
take all possible attacks into account.

It took more than 10 years, but the security of QKD in a noisy
setting was finally solved in a number of papers.
In particular, Shor and Preskill \cite{shor00a} have unified
the earlier proofs by Mayers \cite{mayers96a,mayers01a}
and by Lo and Chau \cite{lo99a}, by using quantum error correction ideas.
[Lo and Chau's proof uses the entanglement distillation approach to
security proof, proposed by Deutsch {\it et al.} \cite{deutsch96}.]
Shor and Preskill showed that BB84 is secure whenever the error rate (commonly
called quantum bit error rate, QBER) is less than 11 percent.
Allowing two-way classical communications between Alice and Bob,
Gottesman and Lo \cite{gottesman03a} have shown that BB84 is secure
whenever the QBER is less than 18.9 percent. Subsequently, Chau \cite{chau02a} extended
the secure region up to 20.0 percent.
An upper bound on the tolerable QBER is also known:
BB84 is known to
be insecure when observed correlations contain no quantum
correlations anymore \cite{curty05a}, which happens when the
average QBER is above 25 percent. \cite{brassard00a} A major
open question is the following: What is the threshold value of
QBER above which BB84 is insecure? Is there really
a gap between the 20\% and the 25\%?

\section{BB84 with practical source in noisy and lossy environment}

Real-life QKD systems suffer from many type of imperfections.
While single photon sources may well be very useful for quantum
computing, it is important to note that single photon sources
are {\it not} needed for QKD. This is good news because
currently single photon sources are rather impractical for QKD.

(a) Source: It is rather common to use attenuated laser pulses as signals.
Those attentuated laser pulses, when phase randomized, follow
a Poissonian distribution
in the number of photons. i.e., the
probability of having $n$ photons in a signal
is given by $P_\mu (n) = e^{- \mu} \mu^n / n\! $ where $\mu$,
chosen by the sender, Alice, is the average number of photons.

For instance, if we use $\mu = 0.1$, then most of the pulses contain no
photons, some contain single photons and a fraction of order 0.005 signals
contains several photons.

(b) Channel: A quantum channel, e.g. an optical fiber or open air, is
lossy as well as noisy.

(c) Detector: Detectors often suffer false detection events due to background and
so-called intrinsic dark counts. Moreover, some misalignment in the
detection system is inevitable.

Let us consider what happens when we use attenuated laser pulses,
rather than perfect single photons, as the source in BB84. \cite{nl00a,brassard00a}
The vacuum component of the signal reduces the signal rate since
no signal will be detected by Bob. The single photon component of
the signal works ideally. The problematic part are the multi-photon signals.
Essentially, each multi-photon signal contains more than one copy of
the polarization information, thus allowing Eve to steal a copy of the information
without Alice and Bob knowing it. More concretely,
the presence of multi-photon signals allows Eve to perform the
{\it photon-number-splitting} (PNS)
attack. In the PNS attack, Eve performs a quantum non-demolition
measurement of the number of photons on each signal emitted by Alice.
Such a measurement tells Eve exactly the number of photons in
a signal without disturbing its polarization.
Now, Eve can act on the signal depending on the total number of photons.
If she finds a vacuum signal, she can resend it to Bob without
introducing any additional errors.
If she finds a multi-photon signal, she splits off one photon and keeps
it in her quantum memory and sends the remainder to Bob.
Note that when Eve sends the reminder to Bob, she may replace
the original lossy quantum channel by a lossless channel.
In other words, Eve may effectively introduce photon-number-dependent loss
in the channel.
Eve's splitting action
does not disturb the signal polarization either in the photon she
splits off, nor in the photon she sends on. Later in the protocol
Alice will reveal the polarization basis of the signal. This will allow
Eve to perform the correct measurement on the single photon she split off,
thereby obtaining perfect information about the polarization of Alice's signal.

The remaining signals are single-photon signals.
Recall that in the PNS attack, Eve has enhanced the transmittance of
multi-photon signals by replacing the original lossy quantum
channel by a lossless one. Therefore,
in order to match the effects of the original loss in the channel, Eve
has to suppress some of the single photon signals. That is to say she has to send
a {\em neutral} signal to Bob that will cause no errors and look like loss.
Of course, the vacuum signal does this job here. The exact rate of the
suppressed signals depends on the mean photon number of the source and the
loss in the channel.

On those single-photon signals that she does not block, Eve
may perform any coherent eavesdropping attack.
This means that in the worst case scenario, all the errors arise from
eavesdropping in single-photon signals.

\begin{figure}[htb]
\includegraphics[width=8cm]{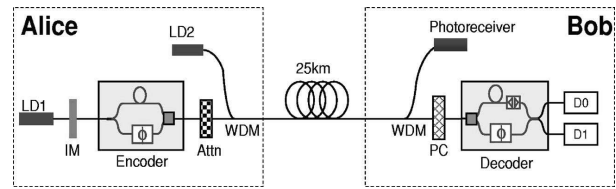}%
\caption{\label{fig:Yuan}Schematics of actual
QKD experiments with phase-encoding of signals. The figure is from
\cite{yuan07} (Courtesy of A. Shields).}
\end{figure}

In the recent years, several groups developed experimental demonstrations of
QKD using imperfect devices. QKD experiments have been
successfully performed over about 100km of commercial Telecom fibers and
also about 100km of open-air. There have even been serious
proposals for performing satellite to ground
QKD experiments, thus enabling a global
quantum cryptographic network through trusted
satellites. One fiber-based QKD set-up by the Cambridge group is
shown in Figure \ref{fig:Yuan}
 Fiber-based QKD systems have
matured over the recent years so that at
least two firms, id Quantique and MagiQ, manufacture them in a commercial setting.

It is important to notice that a security proof that takes into
account all the above imperfections has been given by
Gottesman-Lo-L\"utkenhaus-Preskill (GLLP) \cite{gottesman04a}, building
on earlier work by Inamori,
L\"utkenhaus and Mayers \cite{inamori07a}. As we pointed out
before, it is not "more secure" to
use single-photon sources. In fact, at present, it is much more practical to use
laser pulses, rather than single-photon sources.
Another important issue to mention here is that security
proofs assume models for sending and receiving devices (though not for the
quantum channel). One has therefore always to check that these models fit
the real physical devices.

\section{New Methods}

The rough behaviour of the signal rate according to GLLP is easily
understood. For low and intermediate losses, one can ignore the effects of
errors, and the secret key rate can be understood in terms of the
multi-photon probability of the source, $p_{multi}$, and the observed
probability that Bob receives a signal, $p_{rec}$.
\begin{equation}
G \sim p_{rec} - p_{multi}
\end{equation}
Assuming a Poissonian photon number distribution for the source with mean
photon number $\mu$, we find $p_{multi} =1-(1+\mu)\exp(-\mu)$, and, assuming
that we {\em observe} a probability $p_{rec}$ as in a standard optical
transmission with single photon transmittivity $\eta$ such that $p_{rec}= 1
- \mu \eta \exp(-\mu \eta)$, we can perform a simple optimization over $\mu$
and find the choice $\mu_{opt} \sim \eta$. Therefore, we find
\begin{equation}
G \sim \eta^2 \: .
\end{equation}
This key generation is rather low compared to the communication demand on
optical fiber networks. Moreover, due to the detector imperfections, at some
point the dark counts of the detectors kick in so that we find an effective
cut-off at distances around 20-40 km. In summary,
standard implementations of QKD are limited in distances and key
generation rates.
There are several approaches to solving these problems.

\subsection{Decoy state QKD}
\begin{figure}[htb]
\includegraphics[width=7cm]{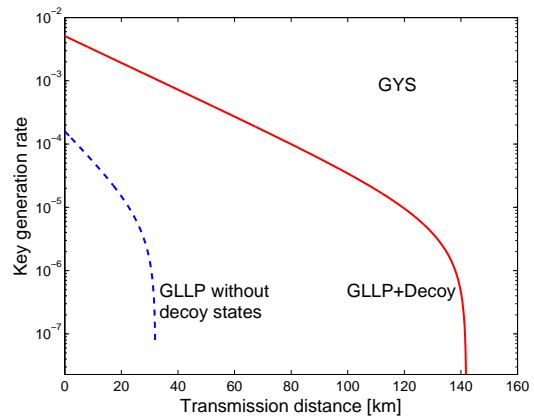}% Here is how to import EPS art
\caption{\label{fig:decoy} Key rates for the experimental set-up in \cite{gobby04a} using the GLLP results
\cite{gottesman04a} with and without the use of decoy states.}
\end{figure}

The first and simplest approach is decoy state QKD \cite{hwang03a,lo05a,wang04suba}
In addition to signal states of average photon number $\mu$,
Alice also creates decoy states of various mean photon numbers $\nu_1$, $\nu_2$, etc.
For instance, Alice may use a variable attenuator to modulate the
intensity of each signal. Consequently, each signal is chosen randomly to be
either a signal state or a decoy state.
Given an $n$-photon signal, an eavesdropper has no idea whether
it comes from a signal state or a decoy state.
Therefore, any attempt for an eavesdropper to suppress single-photon
signals in the signal state will lead also to a suppression of single-photon
signals in the decoy states.
After Bob's acknowledgement of his detection of signals,
Alice broadcasts which signals are signal states and which signals
are decoy states and what types.
By computing the gain (i.e., the ratio of
the number of
detection events to the number of signals sent by Alice) and the
QBER of the decoy state, Alice and Bob will almost surely discover
such a suppression and catch Eve's eavesdropping attack.
As shown by Lo et al. \cite{lo05a}, in the limit of infinite number of choices of
intensities of the decoy states, the only eavesdropping strategy that will
produce the correct gain and QBER for all secretly chosen average
photon number is a standard beam-splitter attack.
As a result, decoy state QKD allows a dramatically higher key
generation rate, $R= O(\eta)$, compared to $R=O(\eta^2)$ for non-decoy protocols
as well as a much higher distance for unconditionally secure QKD
with a practical QKD system. See Figure \ref{fig:decoy}.

How many decoy states are needed? It has been shown that only one or two
type of decoy states are needed for practical protocols. \cite{wang04suba,ma05a}
The first experimental demonstrations of decoy state QKD
has been done. \cite{zhao06a} Given its simplicity,
we expect decoy state QKD to become
a standard technique in the field. Indeed, many follow-up
experiments have now been performed.

\subsection{QKD with strong reference pulses}

The second approach is based on the strong reference pulses idea, dating
back to Bennett's 1992 paper. \cite{bennett92b} The idea is the following.
In addition to
a phase modulated weak signal pulse, Alice sends also a strong unmodulated
reference pulse to Bob through a quantum channel. Quantum information
is encoded in the relative phase between the two pulses---the strong reference pulse
and the signal pulse. Bob splits off a small part of the strong
reference pulse and uses it to interfere with the signal pulse to
measure the relative phase between the two.
In addition, Bob monitors the intensity of the remainder of each
strong reference pulse. Since the reference pulse is strong,
such monitoring can be done easily by, for instance,
a power-meter.

The idea behind this protocol is to make sure that Eve does not have a
neutral signal at her hand which would allow her to suppress signals at will
without causing an error rate. (Such a neutral signal plays a crucial role
in the PNS attack in addition to the multi-photon signals!) Suppose
Eve performs some attacks such as the PNS attack on the signal
pulse. For some measurement outcomes, she would like to selectively
suppress the signal pulse. But how could she do that?

If Eve significantly suppresses the strong
reference pulse, Bob's intensity measurement of the reminder
of the strong reference pulse will find a substantially lower value
than expected. On the other hand, if Eve does not significantly
suppress the strong reference pulse but only suppresses the signal
pulse, then when Bob measures the relative phase between
the strong reference pulse and the signal pulse, he
will find a random outcome (for all conclusive events). So either way, Eve is in trouble.

In a number of recent papers, it has been proven rigorously that
QKD with strong reference pulses can achieve a key generation
rate $R= O (\eta)$. \cite{koashi04a,tamaki06suba} Nonetheless, those proofs require Bob's detection
system with certain suitable properties (one proof requires
Bob has a local oscillator that has been mode-locked to Alice's strong
reference pulse, another requires Bob has a photon detector that
can distinguish multi-photon signals from single photons).
Therefore, they do not apply to standard "threshold" detectors
that do not distinguish single-photon signals from multi-photons.

\subsection{Differential Phase Shift QKD}

A third approach to increase the performance of QKD devices is the the
differential phase shift (DPS) QKD protocol. \cite{inoue05a} Here one uses a coherent train
of laser pulses where the bit information is encoded into the relative phase
between the pulses. But each pulse belongs therefore to two signals! Though
Eve can split photons off the signal trains, these will remain in
non-orthogonal signal states and therefore reveal not their {\em full}
information to Eve! (A similar effect exists already in the B92 protocol with
a strong reference phase!) In the DPS-QKD protocol Eve is now also hampered
again with the suppression of signal states as such a procedure would
require to break the pulse train - which causes errors. The same holds for a
related scheme using time-bins. \cite{stucki05a}

Experimental implementations of DPS QKD have been performed. \cite{diamanti05a}
At present, a rigorous proof of the unconditional security of
differential phase shift QKD is still missing.

\section{Concluding Remarks}

Owing to space limit, we have not talked much about other QKD implementations,
such as those
based on parametric down conversion sources, nor new detectors
such as
superconducting single-photon detectors (SSPDs) and
transition-edge sensor (TES)
detectors. We have omitted also the emerging field of continuous
variable QKD systems which make use of homodyne or heterodyne detection.

Security of QKD is a very slippery subject and one should
work extremely carefully. Regarding a careful analysis
and the formulation of security,
see \cite{renner05a}.
For necessary and sufficient conditions for security, see \cite{horodecki05a}.

In a popular book, "The Code Book", the author, Simon Singh \cite{singh99a}
proposed that quantum cryptography will be the end point of the
evolution of cryptography with the ultimate victory of
the code-makers. Our view is different.
First, quantum cryptography will complement conventional
cryptography, rather than replacing it entirely.
Second, in order to ensure that a practical QKD system is
secure, it is important to verify that the assumptions
made in the security proofs actually hold in the practical system.
Third, QKD does enjoy a fundamental advantage over conventional
cryptography in the sense that,
after a quantum transmission, unlike conventional cryptography,
there is {\it no} classical transcript left for the transmission!
Therefore, for an eavesdropper to break a QKD system, she has to
possess the required quantum technology right at the time of quantum transmission.
For this reason, a skillful eavesdropper can and should invest heavily in
quantum technology {\it now}, rather than later, to
exploit unexpected loopholes in a practical QKD system.
In summary, there is no substitute for battle-testing.
We need quantum hackers as much as quantum cryptographers.
We live in an exciting time where
the interplay between the theory and practice of quantum cryptography
has just begun. The everlasting warfare between code-makers and code-breakers continues.

We thank critical comments on the earlier version of this paper by
Artur Ekert and Renato Renner.
This research is supported by NSERC, CIAR, CRC Program, MITACS, CFI, OIT,
PREA, CIPI and Perimeter Institute for Theoretical Physics.
Research at Perimeter Institute is supported in part by the Government of
Canada through NSERC and the province of Ontario through MEDT.

%\bibliography{qit_20060813,NL_070129}

\begin{thebibliography}{34}
\expandafter\ifx\csname natexlab\endcsname\relax\def\natexlab#1{#1}\fi
\expandafter\ifx\csname bibnamefont\endcsname\relax
  \def\bibnamefont#1{#1}\fi
\expandafter\ifx\csname bibfnamefont\endcsname\relax
  \def\bibfnamefont#1{#1}\fi
\expandafter\ifx\csname citenamefont\endcsname\relax
  \def\citenamefont#1{#1}\fi
\expandafter\ifx\csname url\endcsname\relax
  \def\url#1{\texttt{#1}}\fi
\expandafter\ifx\csname urlprefix\endcsname\relax\def\urlprefix{URL }\fi
\providecommand{\bibinfo}[2]{#2}
\providecommand{\eprint}[2][]{\url{#2}}

\bibitem[{\citenamefont{Welsh}(1988)}]{welsh88a}
\bibinfo{author}{\bibfnamefont{D.}~\bibnamefont{Welsh}},
  \emph{\bibinfo{title}{Codes and Cryptography}} (\bibinfo{publisher}{Oxford
  University Press}, \bibinfo{year}{1988}).

\bibitem[{\citenamefont{Shor}(1997)}]{shor97a}
\bibinfo{author}{\bibfnamefont{P.~W.} \bibnamefont{Shor}},
  \bibinfo{journal}{SIAM J. Comput.} \textbf{\bibinfo{volume}{26}},
  \bibinfo{pages}{1484} (\bibinfo{year}{1997}).

\bibitem[{\citenamefont{Bennett and Brassard}(1984)}]{bennett84a}
\bibinfo{author}{\bibfnamefont{C.~H.} \bibnamefont{Bennett}} \bibnamefont{and}
  \bibinfo{author}{\bibfnamefont{G.}~\bibnamefont{Brassard}}, in
  \emph{\bibinfo{booktitle}{Proceedings of IEEE International Conference on
  Computers, Systems, and Signal Processing, Bangalore, India}}
  (\bibinfo{publisher}{IEEE}, \bibinfo{address}{New York},
  \bibinfo{year}{1984}), pp. \bibinfo{pages}{175--179}.

\bibitem[{\citenamefont{Wiesner}(1983)}]{wiesner83a}
\bibinfo{author}{\bibfnamefont{S.}~\bibnamefont{Wiesner}},
  \bibinfo{journal}{Sigact News} \textbf{\bibinfo{volume}{15}},
  \bibinfo{pages}{78} (\bibinfo{year}{1983}).


\bibitem[{\citenamefont{Ben-Or et~al.}(2005)\citenamefont{Ben-Or, Horodecki,
  Leung, Mayers, and Oppenheim}}]{benor05a}
\bibinfo{author}{\bibfnamefont{M.}~\bibnamefont{Ben-Or}},
  \bibinfo{author}{\bibfnamefont{M.}~\bibnamefont{Horodecki}},
  \bibinfo{author}{\bibfnamefont{D.~W.} \bibnamefont{Leung}},
  \bibinfo{author}{\bibfnamefont{D.}~\bibnamefont{Mayers}}, \bibnamefont{and}
  \bibinfo{author}{\bibfnamefont{J.}~\bibnamefont{Oppenheim}}, in
  \emph{\bibinfo{booktitle}{Second Theory of Cryptography Conference, TCC 2005,
  Cambridge, MA, USA, February 10-12, 2005.}}, edited by
  \bibinfo{editor}{\bibfnamefont{J.}~\bibnamefont{Kilian}}
  (\bibinfo{publisher}{Springer}, \bibinfo{address}{Berlin},
  \bibinfo{year}{2005}), vol. \bibinfo{volume}{3378} of
  \emph{\bibinfo{series}{Lecture Notes in Computer Science}}, pp.
  \bibinfo{pages}{386--406}.

  \bibitem[{\citenamefont{Renner}(2005)}]{renner05a}
\bibinfo{author}{\bibfnamefont{R.}~\bibnamefont{Renner}}, Ph.D. thesis,
  \bibinfo{school}{ETH Z\"urich} (\bibinfo{year}{2005}).


\bibitem[{\citenamefont{Gisin et~al.}(2002)\citenamefont{Gisin, Ribordy,
  Tittel, and Zbinden}}]{gisin02a}
\bibinfo{author}{\bibfnamefont{N.}~\bibnamefont{Gisin}},
  \bibinfo{author}{\bibfnamefont{G.}~\bibnamefont{Ribordy}},
  \bibinfo{author}{\bibfnamefont{W.}~\bibnamefont{Tittel}}, \bibnamefont{and}
  \bibinfo{author}{\bibfnamefont{H.}~\bibnamefont{Zbinden}},
  \bibinfo{journal}{Rev. Mod. Phys.} \textbf{\bibinfo{volume}{74}},
  \bibinfo{pages}{145} (\bibinfo{year}{2002}).

\bibitem[{\citenamefont{Ekert}(1991)}]{ekert91}
\bibinfo{author}{\bibfnamefont{A.~K.} \bibnamefont{Ekert}},
  \bibinfo{journal}{Phys. Rev. Lett} \textbf{\bibinfo{volume}{67}},
  \bibinfo{pages}{661} (\bibinfo{year}{1991}).

\bibitem[{\citenamefont{Bennett}(1992)}]{bennett92b}
\bibinfo{author}{\bibfnamefont{C.~H.} \bibnamefont{Bennett}},
  \bibinfo{journal}{Phys. Rev. Lett.} \textbf{\bibinfo{volume}{68}},
  \bibinfo{pages}{3121} (\bibinfo{year}{1992}).



\bibitem[{\citenamefont{Shor and Preskill}(2000)}]{shor00a}
\bibinfo{author}{\bibfnamefont{P.~W.} \bibnamefont{Shor}} \bibnamefont{and}
  \bibinfo{author}{\bibfnamefont{J.}~\bibnamefont{Preskill}},
  \bibinfo{journal}{Phys. Rev. Lett.} \textbf{\bibinfo{volume}{85}},
  \bibinfo{pages}{441} (\bibinfo{year}{2000}).

\bibitem[{\citenamefont{Mayers}(1996)}]{mayers96a}
\bibinfo{author}{\bibfnamefont{D.}~\bibnamefont{Mayers}}, in
  \emph{\bibinfo{booktitle}{Advances in Cryptology \,---\, Proceedings of
  Crypto '96}} (\bibinfo{publisher}{Springer}, \bibinfo{address}{Berlin},
  \bibinfo{year}{1996}), pp. \bibinfo{pages}{343--357}.

\bibitem[{\citenamefont{Mayers}(2001)}]{mayers01a}
\bibinfo{author}{\bibfnamefont{D.}~\bibnamefont{Mayers}},
  \bibinfo{journal}{JACM} \textbf{\bibinfo{volume}{48}}, \bibinfo{pages}{351}
  (\bibinfo{year}{2001}).

\bibitem[{\citenamefont{Lo and Chau}(1999)}]{lo99a}
\bibinfo{author}{\bibfnamefont{H.-K.} \bibnamefont{Lo}} \bibnamefont{and}
  \bibinfo{author}{\bibfnamefont{H.~F.} \bibnamefont{Chau}},
  \bibinfo{journal}{Science} \textbf{\bibinfo{volume}{283}},
  \bibinfo{pages}{2050} (\bibinfo{year}{1999}).


\bibitem[{\citenamefont{Deutsch}(1996)}]{deutsch96}
\bibinfo{author}{\bibfnamefont{D.}~\bibnamefont{Deutsch}},
  \bibinfo{author}{\bibfnamefont{A.}~\bibnamefont{Ekert}},
  \bibinfo{author}{\bibfnamefont{R.}~\bibnamefont{Josza}},
\bibinfo{author}{\bibfnamefont{C.}~\bibnamefont{Macchiavello}},
\bibinfo{author}{\bibfnamefont{S.}~\bibnamefont{Popescu}},
  \bibnamefont{and}
  \bibinfo{author}{\bibfnamefont{A.}~\bibnamefont{Sanpera}},
  \bibinfo{journal}{Phys. Rev. Lett.} \textbf{\bibinfo{volume}{77}},
  \bibinfo{pages}{2818} (\bibinfo{year}{1996}).

\bibitem[{\citenamefont{Gottesman and Lo}(2003)}]{gottesman03a}
\bibinfo{author}{\bibfnamefont{D.}~\bibnamefont{Gottesman}} \bibnamefont{and}
  \bibinfo{author}{\bibfnamefont{H.-K.} \bibnamefont{Lo}},
  \bibinfo{journal}{IEEE Trans. Inf. Theory} \textbf{\bibinfo{volume}{49}},
  \bibinfo{pages}{457} (\bibinfo{year}{2003}).

\bibitem[{\citenamefont{Chau}(2002)}]{chau02a}
\bibinfo{author}{\bibfnamefont{H.}~\bibnamefont{Chau}}, \bibinfo{journal}{Phys.
  Rev. A} \textbf{\bibinfo{volume}{66}}, \bibinfo{pages}{60302}
  (\bibinfo{year}{2002}).

\bibitem[{\citenamefont{Curty et~al.}(2005)\citenamefont{Curty, G\"uhne,
  Lewenstein, and L\"utkenhaus}}]{curty05a}
\bibinfo{author}{\bibfnamefont{M.}~\bibnamefont{Curty}},
  \bibinfo{author}{\bibfnamefont{O.}~\bibnamefont{G\"uhne}},
  \bibinfo{author}{\bibfnamefont{M.}~\bibnamefont{Lewenstein}},
  \bibnamefont{and}
  \bibinfo{author}{\bibfnamefont{N.}~\bibnamefont{L\"utkenhaus}},
  \bibinfo{journal}{Phys. Rev. A} \textbf{\bibinfo{volume}{71}},
  \bibinfo{pages}{022306} (\bibinfo{year}{2005}).

\bibitem[{\citenamefont{Brassard et~al.}(2000)\citenamefont{Brassard,
  L\"utkenhaus, Mor, and Sanders}}]{brassard00a}
\bibinfo{author}{\bibfnamefont{G.}~\bibnamefont{Brassard}},
  \bibinfo{author}{\bibfnamefont{N.}~\bibnamefont{L\"utkenhaus}},
  \bibinfo{author}{\bibfnamefont{T.}~\bibnamefont{Mor}}, \bibnamefont{and}
  \bibinfo{author}{\bibfnamefont{B.}~\bibnamefont{Sanders}},
  \bibinfo{journal}{Phys. Rev. Lett.} \textbf{\bibinfo{volume}{85}},
  \bibinfo{pages}{1330} (\bibinfo{year}{2000}).

\bibitem[{\citenamefont{L\"utkenhaus}(2000)}]{nl00a}
\bibinfo{author}{\bibfnamefont{N.}~\bibnamefont{L\"utkenhaus}},
  \bibinfo{journal}{Phys. Rev. A} \textbf{\bibinfo{volume}{61}},
  \bibinfo{pages}{052304} (\bibinfo{year}{2000}).

\bibitem[{\citenamefont{Yuan et~al.}(2007)\citenamefont{Yuan, Sharpe, and
  Shields}}]{yuan07}
\bibinfo{author}{\bibfnamefont{Z.L.}~\bibnamefont{Yuan}},
\bibinfo{author}{\bibfnamefont{A.W.}~\bibnamefont{Sharpe}},
\bibnamefont{and}
  \bibinfo{author}{\bibfnamefont{A.J.}~\bibnamefont{Shields}},
  \bibinfo{journal}{Appl. Phys. Lett.} \textbf{\bibinfo{volume}{90}},
  \bibinfo{pages}{011118} (\bibinfo{year}{2007}).

\bibitem[{\citenamefont{Gottesman et~al.}(2004)\citenamefont{Gottesman, Lo,
  L\"utkenhaus, and Preskill}}]{gottesman04a}
\bibinfo{author}{\bibfnamefont{D.}~\bibnamefont{Gottesman}},
  \bibinfo{author}{\bibfnamefont{H.-K.} \bibnamefont{Lo}},
  \bibinfo{author}{\bibfnamefont{N.}~\bibnamefont{L\"utkenhaus}},
  \bibnamefont{and} \bibinfo{author}{\bibfnamefont{J.}~\bibnamefont{Preskill}},
  \bibinfo{journal}{Quant. Inf. Comp.} \textbf{\bibinfo{volume}{4}},
  \bibinfo{pages}{325} (\bibinfo{year}{2004}).


\bibitem[{\citenamefont{Inamori et~al.}(2007)\citenamefont{Inamori,
  L\"utkenhaus, and Mayers}}]{inamori07a}
\bibinfo{author}{\bibfnamefont{H.}~\bibnamefont{Inamori}},
  \bibinfo{author}{\bibfnamefont{N.}~\bibnamefont{L\"utkenhaus}},
  \bibnamefont{and} \bibinfo{author}{\bibfnamefont{D.}~\bibnamefont{Mayers}},
  \bibinfo{journal}{Eur. Phys. J. D}  (\bibinfo{year}{2007}); \bibinfo{howpublished}{quant-ph/0107017}.

\bibitem[{\citenamefont{Hwang}(2003)}]{hwang03a}
\bibinfo{author}{\bibfnamefont{W.-Y.} \bibnamefont{Hwang}},
  \bibinfo{journal}{Phys. Rev. Lett} \textbf{\bibinfo{volume}{91}},
  \bibinfo{pages}{57901} (\bibinfo{year}{2003}).

\bibitem[{\citenamefont{Lo et~al.}(2005)\citenamefont{Lo, Ma, and
  Chen}}]{lo05a}
\bibinfo{author}{\bibfnamefont{H.-K.} \bibnamefont{Lo}},
  \bibinfo{author}{\bibfnamefont{X.}~\bibnamefont{Ma}}, \bibnamefont{and}
  \bibinfo{author}{\bibfnamefont{K.}~\bibnamefont{Chen}},
  \bibinfo{journal}{Phys. Rev. Lett.} \textbf{\bibinfo{volume}{94}},
  \bibinfo{pages}{230504} (\bibinfo{year}{2005}).

\bibitem[{\citenamefont{Wang}()}]{wang04suba}
\bibinfo{author}{\bibfnamefont{X.-B.} \bibnamefont{Wang}},
   \bibinfo{journal}{Phys. Rev. Lett.} \textbf{\bibinfo{volume}{94}},
  \bibinfo{pages}{230503} (\bibinfo{year}{2005}).

\bibitem[{\citenamefont{Gobby et~al.}(2004)\citenamefont{Gobby, Yuan, and
  Shields}}]{gobby04a}
\bibinfo{author}{\bibfnamefont{C.}~\bibnamefont{Gobby}},
  \bibinfo{author}{\bibfnamefont{Z.}~\bibnamefont{Yuan}}, \bibnamefont{and}
  \bibinfo{author}{\bibfnamefont{A.}~\bibnamefont{Shields}},
  \bibinfo{journal}{Appl. Phys. Lett.} \textbf{\bibinfo{volume}{84}},
  \bibinfo{pages}{3762} (\bibinfo{year}{2004}).


\bibitem[{\citenamefont{Ma et~al.}(2005)\citenamefont{Ma, Zhao, and
  Lo}}]{ma05a}
\bibinfo{author}{\bibfnamefont{B.}~\bibnamefont{Ma}, \bibfnamefont{X.and~Qi}},
  \bibinfo{author}{\bibfnamefont{Y.}~\bibnamefont{Zhao}}, \bibnamefont{and}
  \bibinfo{author}{\bibfnamefont{H.-K.} \bibnamefont{Lo}},
  \bibinfo{journal}{Phys. Rev. A} \textbf{\bibinfo{volume}{72}},
  \bibinfo{pages}{012326} (\bibinfo{year}{2005}).

\bibitem[{\citenamefont{Zhao et~al.}(2006)\citenamefont{Zhao, Qi, Ma, Lo, and
  Qian}}]{zhao06a}
\bibinfo{author}{\bibfnamefont{Y.}~\bibnamefont{Zhao}},
  \bibinfo{author}{\bibfnamefont{B.}~\bibnamefont{Qi}},
  \bibinfo{author}{\bibfnamefont{X.}~\bibnamefont{Ma}},
  \bibinfo{author}{\bibfnamefont{H.-K.} \bibnamefont{Lo}}, \bibnamefont{and}
  \bibinfo{author}{\bibfnamefont{L.}~\bibnamefont{Qian}},
  \bibinfo{journal}{Phys. Rev. Lett.} \textbf{\bibinfo{volume}{96}},
  \bibinfo{pages}{070502} (\bibinfo{year}{2006}).

\bibitem[{\citenamefont{Koashi}(2004)}]{koashi04a}
\bibinfo{author}{\bibfnamefont{M.}~\bibnamefont{Koashi}},
  \bibinfo{journal}{Physical Review Letters} \textbf{\bibinfo{volume}{93}},
  \bibinfo{eid}{120501} (pages~\bibinfo{numpages}{4}) (\bibinfo{year}{2004}).

\bibitem[{\citenamefont{Tamaki et~al.}()\citenamefont{Tamaki, L\"utkenhaus,
  Koashi, and Batuwantudawe}}]{tamaki06suba}
\bibinfo{author}{\bibfnamefont{K.}~\bibnamefont{Tamaki}},
  \bibinfo{author}{\bibfnamefont{N.}~\bibnamefont{L\"utkenhaus}},
  \bibinfo{author}{\bibfnamefont{M.}~\bibnamefont{Koashi}}, \bibnamefont{and}
  \bibinfo{author}{\bibfnamefont{J.}~\bibnamefont{Batuwantudawe}},
  \bibinfo{howpublished}{quant-ph/0607082}.

\bibitem[{\citenamefont{Inoue and Honjo}(2005)}]{inoue05a}
\bibinfo{author}{\bibfnamefont{K.}~\bibnamefont{Inoue}} \bibnamefont{and}
  \bibinfo{author}{\bibfnamefont{T.}~\bibnamefont{Honjo}},
  \bibinfo{journal}{Phys. Rev. A} \textbf{\bibinfo{volume}{71}},
  \bibinfo{pages}{042305} (\bibinfo{year}{2005}).

\bibitem[{\citenamefont{Stucki et~al.}(2005)\citenamefont{Stucki, Brunner,
  Gisin, Scarani, and Zbinden}}]{stucki05a}
\bibinfo{author}{\bibfnamefont{D.}~\bibnamefont{Stucki}},
  \bibinfo{author}{\bibfnamefont{N.}~\bibnamefont{Brunner}},
  \bibinfo{author}{\bibfnamefont{N.}~\bibnamefont{Gisin}},
  \bibinfo{author}{\bibfnamefont{V.}~\bibnamefont{Scarani}}, \bibnamefont{and}
  \bibinfo{author}{\bibfnamefont{H.}~\bibnamefont{Zbinden}},
  \bibinfo{journal}{Appl. Phys. Lett.} \textbf{\bibinfo{volume}{87}},
  \bibinfo{pages}{194108} (\bibinfo{year}{2005}).

\bibitem[{\citenamefont{Diamanti et~al.}(2005)\citenamefont{Diamanti, Takesue,
  Honjo, Inoue, and Yamamoto}}]{diamanti05a}
\bibinfo{author}{\bibfnamefont{E.}~\bibnamefont{Diamanti}},
  \bibinfo{author}{\bibfnamefont{H.}~\bibnamefont{Takesue}},
  \bibinfo{author}{\bibfnamefont{T.}~\bibnamefont{Honjo}},
  \bibinfo{author}{\bibfnamefont{K.}~\bibnamefont{Inoue}}, \bibnamefont{and}
  \bibinfo{author}{\bibfnamefont{Y.}~\bibnamefont{Yamamoto}},
  \bibinfo{journal}{Phys. Rev. A} \textbf{\bibinfo{volume}{72}},
  \bibinfo{pages}{052311} (\bibinfo{year}{2005}).


\bibitem[{\citenamefont{Horodecki et~al.}(2005)\citenamefont{Horodecki,
  Horodecki, Horodecki, and Oppenheim}}]{horodecki05a}
\bibinfo{author}{\bibfnamefont{K.}~\bibnamefont{Horodecki}},
  \bibinfo{author}{\bibfnamefont{M.}~\bibnamefont{Horodecki}},
  \bibinfo{author}{\bibfnamefont{P.}~\bibnamefont{Horodecki}},
  \bibnamefont{and}
  \bibinfo{author}{\bibfnamefont{J.}~\bibnamefont{Oppenheim}},
  \bibinfo{journal}{Phys. Rev. Lett.} \textbf{\bibinfo{volume}{94}},
  \bibinfo{pages}{160502} (\bibinfo{year}{2005}).

\bibitem[{\citenamefont{Singh}(1999)}]{singh99a}
\bibinfo{author}{\bibfnamefont{S.}~\bibnamefont{Singh}},
  \emph{\bibinfo{title}{The Code Book: The Science of Secrecy from Ancient
  Egypt to Quantum Cryptography}} (\bibinfo{publisher}{Doubleday},
  \bibinfo{address}{New York}, \bibinfo{year}{1999}).

\end{thebibliography}

\end{document}